\documentclass[]{aa}  

\usepackage{natbib}
\bibpunct{(}{)}{;}{a}{}{,}
\usepackage{graphicx}
\usepackage{revsymb}	
\usepackage{amsmath}	
\usepackage[usenames]{color}	
\usepackage{epstopdf}	
\usepackage{txfonts}
\usepackage{amssymb}
\usepackage{color}
\usepackage[justification=centering]{caption}
\usepackage{geometry} 
\usepackage[parfill]{parskip} 
\usepackage{amssymb}
\usepackage{slantsc}
\usepackage{array}
\usepackage{url}
\usepackage{amsfonts}
\usepackage[colorlinks=true,linkcolor=black, urlcolor=black, citecolor=black]{hyperref}
\usepackage{hyperref} 
\usepackage{sidecap}
\usepackage{graphicx}
\usepackage{xcolor}
\usepackage{nicefrac}
\usepackage{subfigure}
\usepackage{epsfig}
\colorlet{rouge}{red!70!darkgray}

\begin{document}
\title{Analysis of the linear approximation of seismic inversions for various structural pairs}
\author{G. Buldgen\inst{1}\and D. R. Reese\inst{2}\and M. A. Dupret\inst{1}}
\institute{Institut d’Astrophysique et Géophysique de l’Université de Liège, Allée du 6 août 17, 4000 Liège, Belgium \and LESIA, Observatoire de Paris, PSL Research University, CNRS, Sorbonne Universités, UPMC Univ. Paris 06, Univ. Paris Diderot, Sorbonne Paris Cité, 5 place Jules Janssen, 92195 Meudon Cedex, France}
\date{April, 2015}
\abstract{Thanks to the space-based photometry missions CoRoT and Kepler, we now benefit from a wealth of seismic data for stars other  than the sun. In the future, K2, Tess, and Plato will complement this data and provide observations in addition to those already at hand. The availability of this data leads to questions on how it is feasible to extend kernel-based, linear structural inversion techniques to stars other than the sun. Linked to the inversion problem is the question of the validity of the linear assumption. In this study, we analyse the limitations of this assumption with respect to changes of structural variables.}
{We wish to provide a more extended theoretical background to structural linear inversions by doing a study of the validity of the linear assumption for various structural variables. We thus point towards limitations in inversion techniques in the asteroseismic and helioseismic cases.}
{First, we recall the origins of the linear assumption for structural stellar inversions and explain its importance for asteroseismic studies. We also briefly recall the impact of unknown structural quantities such as the mass and the radius of the star on structural inversion results. We then explain how kernels for new structural variables can be derived using two methods, one suited to asteroseismic targets, the other to helioseismic targets. For this second method, we present a new structural pair, namely the $(A,Y)$ structural kernels. The kernels are then tested in various numerical experiments that enable us to evaluate the weaknesses of different structural pairs and the domains of validity of their respective linear regime.}
{The numerical tests we carry out allow us to disentangle the impact of various uncertainties in stellar models on the verification of the linear integral relations. We show the importance of metallicity, the impact of the equation of state, extra-mixing, and inaccuracies in the microphysics on the verification of these relations. We also discuss the limitations of the method of conjugated functions due to the lack of extremely precise determinations of masses or radii in the asteroseismic context.}{}
\maketitle
\section{Introduction} \label{SecIntro}
Asteroseismology is now considered the golden path to the study of stellar structure. This young research field benefits from high quality data for a large sample of stars stemming from the successes of the CoRoT, Kepler, and K2 missions \citep{Baglin, Borucki, Chaplin}. More specifically, the detection of solar-like oscillations in a large sample of stars now allows a more accurate study of stellar structure. In the future, the Tess and Plato missions \citep{Rauer} will carry on what is now called the space-photometry revolution. 
\\
\\
Historically, the successes of asteroseismology were preceded by the successes of helioseismology, the study of solar pulsations. Indeed, the quality of seismic data of the sun is still far beyond what is achievable for other stars, even in the era of the space missions. In the particular field of helioseismology, numerous methods were developed to obtain constraints on the solar structure. Amongst them, inversion techniques lead to the successful determination of the position of the base of the convective envelope, the helium abundance in this region, and the rotational profile of the sun \citep{Schou}. The determination of the sound speed and density profiles also demonstrated the importance of microscopic diffusion for solar models \citep[e.g.][]{BasuReview}.
\\
\\
In the context of asteroseismology, the use of these inversion techniques can now be considered for a limited number of targets under the conditions of validity of all the hypotheses hiding behind the basic equations defining their applicability domain. The most constraining of these hypotheses is to assume a linear relation between frequency differences and structural differences. In this paper, we propose to analyse in depth the issues surrounding the linearity of these relations for various structural pairs and more specifically for kernels of the convective parameter. To that extent, we derive new structural kernels for the convective parameter and helium abundance and compare their linear behaviour to other, pre-existing kernels. Differences in the verification of the linearity of the frequency-structure relation mean that care should be taken when combining seismic diagnostics of various kernels, even in the solar case. Differences can stem from the intrinsic non-linear behaviour of the structural variables considered, but can also be the results of inaccuracies in terms of numerical quality of the models and/or kernels.
\section{The variational principle and linear frequency-structure equations} \label{SecVariaIntro}
The variational principle is a well-known property of adiabatic stellar oscillation equations. In fact, it can be extended to more general objects than stars and generalized beyond the classical case presented in helio- and asteroseismology. The history of the variational principle can be traced back to stability analysis in structural mechanics, but its application in seismology stems from the pioneering work of Chandrasekhar \citep{Chandra} and the generalization of his study by other authors the following years \citep{Clement, LyndenBell}.

Far beyond the historical interest of the discovery of this mathematical property, the hypotheses that lay behind this principle are still important since they are at the heart of intrinsic limitations of the frequency-structure relation. From the mathematical point of view, the variational principle is a consequence of the symmetry of the operator associated with adiabatic stellar pulsations. Mathematically, this means that given two functions $\vec{\xi}$ and $\vec{\psi}$ and $\mathcal{D}$ the operator associated with adiabatic oscillations, we have the property
\begin{align}
<\xi,\mathcal{D}(\psi)>=<\psi,\mathcal{D}(\xi)>,
\end{align}
 where $<,>$ denotes the scalar product associated with the functional space defined as
\begin{align}
<\vec{\xi}, \vec{\psi}> = \int_{V} \rho \vec{\xi} . \vec{\psi}^{*} dV,
\end{align}
with the symbol $^{*}$ denoting the complex conjugate. The absence of symmetry in the non-adiabatic case is the reason why all inversions are carried out using the hypothesis of adiabaticity of stellar oscillations.

Moreover, the hypothesis of linearity of the frequency-structure relations is a strong restriction to the validity of seismic inversions and in this section we briefly discuss how this hypothesis influences structural diagnostics from inversion techniques. The relation between perturbations of the frequencies and small perturbations of the stellar structure can be obtained by perturbing the variational expression of the pulsation frequencies. The symmetry of the stellar pulsation operator is then used to eliminate perturbations of the eigenfunctions in the resultant expression. Other effects such as perturbations of the eigenfunctions associated with each pulsation frequency can be neglected to first order. This implies that a direct relation can be obtained between structural differences and pulsation frequencies only. This relation is formally written:
\begin{align}
\delta \nu=\frac{<\vec{\xi},\delta \mathcal{D}(\vec{\xi})>}{<\vec{\xi}, \vec{\xi}>}, \label{EqVar}
\end{align}
with $\delta \nu$ the perturbation of an oscillation frequency, $\delta \mathcal{D}$ the associated perturbation to the operator of adiabatic oscillations. The eigenfunctions $\vec{\xi}$ and the unperturbed operator $\mathcal{D}$ are known and defined from the reference model.

In practice, Eq. (\ref{EqVar}) implies that small differences in frequencies can be used to analyse the associated differences in the operator of adiabatic pulsations. The main problem is the scalar product which implies integral relations and thus an ill-posed problem. However, it should be noted that the validity of the variational expression is limited, since we are speaking of small perturbations, the term ``small'' being misleading because it is often retroactively defined. In other words, perturbations are small because the variational expression is satisfied, but the quantification of how small a perturbation can be and if all variables can be similarly perturbed remains uncertain.

The classical equation for inversion techniques is the result of further developments introduced in the variational expression, assuming spherical symmetry of the star, and carrying out integration by parts and permutation of integrals. This ends leading to the following formally simple equation \citep[See][for a full demonstration of this expression and its hypotheses.]{Gough}:
\begin{align}
\frac{\delta \nu^{n,l}}{\nu^{n,l}}=\int_{0}^{R}K^{n,l}_{\rho,c^{2}}\frac{\delta \rho}{\rho}dr+\int_{0}^{R}K^{n,l}_{c^{2},\rho}\frac{\delta c^{2}}{c^{2}}dr+\mathcal{O}(\delta^{2}), \label{eqrhoc2}
\end{align}
with the following definition:
\begin{align}
\frac{\delta x}{x}=\frac{x_{obs}-x_{ref}}{x_{ref}}.
\end{align}
The quantity $x$ can be the oscillation frequency of a particular mode, $\nu^{n,l}$, the density, $\rho$, the squared adiabatic sound speed, $c^{2}$, or other quantities for which kernels can be derived.

First of all, we note that to this expression is usually added the surface effects term, which is an empirical correction that is added to Eq. (\ref{eqrhoc2}) to take into account the improper modelling of surface layers in the computation of oscillation frequencies of stellar models. In this study, we do not consider this surface term since we will only compare theoretical models and the validity of the linear approximation for various test cases between these models.

In terms of seismic diagnostic, the linear hypothesis puts strong restrictions on the applicability of inversion techniques. In fact, in some regions of the Hertzsprung-Russell diagram, it seems obvious that differences in frequencies may not be linearly related to structural changes. For example, in evolved stars, changes in the mixed modes frequencies will have strong impact on the coupling of the p and g mode cavities. Thus a small change in frequency will imply a strong change in the eigenfunctions. In this particular case, the second order terms neglected in the variational analysis may well become dominant and have to be modelled to efficiently use kernel based inversions as a seismic diagnostic.

For the case of p modes observed in solar-like stars, one could say that provided the model is good enough, the linear approximation may be used. However, the linear approximation as presented is usually for a slow-rotating, non-magnetic, isolated star\footnote{By isolated, we mean that it is not in a close binary system where the gravitational influence of the neighbouring companion would change the geometry of the star and its oscillation modes}. The problem is not to carry out the inversion, since that can be done provided a sufficient number of frequencies is available, but to decide whether the inverted results can be trusted. The errors due to linearity are intrinsically not seen by the inversion technique. However, it is still possible to witness their effects on inversion results and indirectly assess the quality of the reference model. To do so, one simply has to start from various reference models and analyse the variation of the results with the model. This simple and straightforward method is well-adapted to global optimization techniques which generate a large sample of models. However, this does not mean that using a large number of models, one can go beyond the linear approximation of the variational principle, it only implies that one can analyse the errors coming from the non-linear effects and decide whether the results should be trusted or not.

An other important aspect of asteroseismic inversions which has been reported by \citet{BasuSca} and described in \citet{Buldgentu} is that the inversion scales its results implicitly. This scaling stems from the assumption that integral relations are defined on the same domain for the reference model as for the observed target. In other words, if we define $R_{Ref}$, the radius of the reference model and $R_{Tar}$, the radius of the observed target, the inversion will wrongly consider that both radii are equal. However, since the inversion uses seismic information, the mean density of the observed target is known. Consequently, the mass of the scaled target, denoted here $\tilde{M}_{Tar}$, which is studied by the inversion is $\tilde{M}_{Tar}=R^{3}_{Ref}\bar{\rho}_{Tar}$, with $\bar{\rho}_{Tar}$ the mean density of the observed target. This means that structural quantities such as the squared sound speed, or indicators defined by integrated quantities, are not determined for the observed target itself, but for the scaled one and are related to the observed quantities through an homology. While this does not reduce the diagnostic potential of inversion techniques, it should of course be taken into account when comparing results inverted from various reference models.
\section{Changing the structural pair} \label{SecStrucChange}
The calculation of new kernels is particularly interesting in the context of asteroseismology, where the change of structural variables can significantly improve the ability to fit a certain target while reducing the contribution from the so-called ``cross-term". Additional kernels have also been used in helioseismology to test the equation of state used in solar models and to try to determine the helium abundance in the convective envelope. In this section, we present two methods to derive additional structural kernels from Eq. \ref{eqrhoc2} and discuss in more details their implementation and respective strengths and weaknesses.
\subsection{Direct method} \label{SecMethodDirect}
We call this approach direct because it consists in a direct change of variables within Eq. \ref{eqrhoc2} (or any similar relation), from which a linear differential equation is derived (This equation can be of the first, second or third order depending on the variables involved). The resolution of this equation allows us to determine new kernels, provided the proper boundary conditions are applied. 
\\
\\
In practice, this method gives access to any function of $\rho$, $c^{2}$, $\Gamma_{1}$ or their integrals (e.g. hydrostatic pressure, $P$ or the gravitational acceleration, $g$), or combinations of these variables (e.g. the squared isothermal sound speed, $u=\frac{P}{\rho}$)\footnote{More generally, this function could be written $f(\rho,P,g,\Gamma_{1})$ or $f(\rho,P,g,Y)$ with any $f$ that can be written in terms of linear perturbations of these quantities.}. However, it should be noted that this method does not give access to any function of the derivative of the density without further integration by parts when deriving the differential equation. The kernels that can be obtained through the direct method are listed in Table \ref{TabPairs}. We mention that this list contains only kernels for which the equations have been derived, but one could be interested to define new thermodynamical variables and to obtain kernels for these new variables.
\begin{table}[t]
\caption{Summary of the properties of the differential equations for various structural pairs with the direct method.}
\label{TabPairs}
  \centering
\begin{tabular}{r | c | c}
\hline \hline
 \textbf{Pair}& \textbf{Order of o.d.e.}& \textbf{Integration by parts} \\ \hline
 $\rho,\Gamma_{1}$ (or $Y$) & $0$-algebraic & No \\
$g,\Gamma_{1}$ (or $Y$) & $1$ & No \\ 
$P,\Gamma_{1}$ (or $Y$) & $2$ & No \\ 
$c^{2},\Gamma_{1}$ (or $Y$) & $2$ & No \\
$u,\Gamma_{1}$ (or $Y$) & $2$ & No \\
$A,\Gamma_{1}$ (or $Y$) & $3$ & Yes \\
$N^{2},c^ {2}$ & $3$ & Yes\\
\hline
\end{tabular}
\end{table}
\\
\\
This method has been partially presented in a previous paper \citep{Buldgentu} and referred to as Masters' method, because it was developed as an extension of an approach presented in \citet{Masters} for geophysical applications that was mentioned in \citet{Gough} as a potential method for obtaining kernels for the Brunt-Väisälä frequency\footnote{We describe how this can be done in appendix \ref{SecAYDirect}}. Originally, Masters' approach proposed to solve directly the integral relations between structural kernels used in geophysics. In asteroseismology, the method could have been similar. First, we start with Eq. \ref{eqrhoc2} and consider for example the change from the $(\rho,c^{2})$ structural pair to the $(g,c^{2})$ structural pair, where $g$ is the gravitational acceleration and is written:
\begin{align}
g=\frac{Gm(r)}{r^{2}},
\end{align}
with $m(r)$ the mass of stellar material contained in a sphere of radius $r$ and being defined:
\begin{align}
m(r)=\int_{0}^{r}4 \pi \tilde{r}^{2} \rho d\tilde{r}.
\end{align}
If we consider the linear relative perturbation of the gravity acceleration, we obtain:
\begin{align}
\frac{\delta g}{g}= \frac{\frac{G \delta m}{r^{2}}}{\frac{G m}{r^{2}}}=\frac{\delta m}{m}=\frac{1}{m}\int_{0}^{r}4 \pi \tilde{r}^{2} \delta \rho d\tilde{r}.
\end{align}
This expression can be used directly in the integral relations for the structural kernels. Indeed, if kernels of the pair $(g,c^{2})$ can be found, they must satisfy the following relation:
\begin{align}
\frac{\delta \nu^{n,l}}{\nu^{n,l}}& =\int_{0}^{R}K^{n,l}_{g,c^{2}} \frac{\delta g}{g}dr +\int_{0}^{R}K^{n,l}_{c^{2},g}\frac{\delta c^{2}}{c^{2}}dr \nonumber \\
& = \int_{0}^{R}K^{n,l}_{\rho,c^{2}}\frac{\delta \rho}{\rho}dr+\int_{0}^{R}K^{n,l}_{c^{2},\rho}\frac{\delta c^{2}}{c^{2}}dr \label{eqlinear}.
\end{align}
From the second equality, we have the integral relation that we searched. One has only to introduce the perturbation of the gravitational acceleration and permute the integrals such that the perturbation of density is in the outermost integral. From there it is easy to obtain a simple relation between kernels. Indeed, Eq. \ref{eqlinear} must be satisfied for any perturbation within the linear regime, since the kernels must be dependent on the reference model only. We then obtain simple relations for each kernel:
\begin{align}
K^{n,l}_{c^{2},g}&=K^{n,l}_{c^{2},\rho} \label{eqc2g} \\
4 \pi r^{2} \rho \int_{r}^{R}\frac{K^{n,l}_{g,c^{2}}}{m}&=K^{n,l}_{\rho,c^{2}}. \label{eqgrho}
\end{align}
The proposition of \citet{Masters} was to solve directly this integral relation, which can be done using, for example, an iterative relaxation method to solve the integral equation. In practice, we favour a more efficient approach by deriving a differential equation for these kernels, simply by taking the derivative of Eq. \ref{eqgrho} after having divided it by $r^{2}\rho$. We then obtain the following very simple differential equation:
\begin{align}
-m\frac{d}{dr}\left( \frac{K^{n,l}_{\rho,c^{2}}}{\rho r^{2}} \right)&= K^{n,l}_{g,c^{2}},
\end{align}
Since this equation is extremely simple, the kernels are straightforward to obtain. However, this development was just for the sake of illustration and a good example of the difficulties associated with this method is illustrated in appendix \ref{SecAYDirect}. 
\\
\\
A more elaborated case, which has already been involved in practical applications is that of the $(u,\Gamma_{1})$ and $(u,Y)$ kernels. These kernels are obtained by solving a second order differential equation which is recalled here:
\begin{align}
-y\frac{d^{2}\mathcal{K}^{'}}{(dy)^{2}}&+\left[ \frac{2 \pi y^{3/2}\tilde{\rho}}{\tilde{m}}-3 \right]\frac{d \mathcal{K}^{'}}{dy}=y\frac{d^{2} \mathcal{K}}{(dy)^{2}} \nonumber\\ 
&-\left[ \frac{2 \pi y^{3/2} \tilde{\rho}}{\tilde{m}}-3 +\frac{\tilde{m} \tilde{\rho}}{2 y^{1/2} \tilde{P}} \right]\frac{d \mathcal{K}}{dy} \nonumber\\ &+ \left[ \frac{\tilde{m} \tilde{\rho}}{4 y \tilde{P}^{2}} \frac{d\tilde{P}}{dx} - \frac{\tilde{m}}{4y \tilde{P}} \frac{d \tilde{\rho}}{dx}-\frac{3}{4y^{1/2} \tilde{P}}\frac{d\tilde{P}}{dx} -\frac{\tilde{m} \tilde{\rho}}{2y^{3/2}\tilde{P}}\right] \mathcal{K}, \label{eqdiffugamma}
\end{align}
with $\mathcal{K}=\frac{K^{n,l}_{u,\Gamma_{1}}}{r^{2}\rho}$ and $\mathcal{K}^{'}=\frac{K^{n,l}_{\rho,\Gamma_{1}}}{r^{2}\rho}$ in the case of the differential equation of the $(u,\Gamma_{1})$ kernels or with with $\mathcal{K}=\frac{K^{n,l}_{u,Y}}{r^{2}\rho}$ and $\mathcal{K}^{'}=\frac{K^{n,l}_{\rho,Y}}{r^{2}\rho}$ for the equation of the $(u,Y)$ kernels. Furthermore, in Eq. \ref{eqdiffugamma}, one has the following definitions: $x=\frac{r}{R}$, $y=x^{2}$, $\tilde{m}=\frac{m}{M}$, $\tilde{\rho}=\frac{R^{3}\rho}{M}$, $\tilde{P}=\frac{R^{4}P}{GM}$. We also recall here that using kernels such as the $(\rho,Y)$ or the $(u,Y)$ kernels requires to introduce the equation of state by using the following definition:
\begin{align}
\frac{\delta \Gamma_{1}}{\Gamma_{1}} &= \left(\frac{\partial \ln \Gamma_{1}}{\partial \ln P}\right)_{Z,Y,\rho} \frac{\delta P}{P} + \left(\frac{\partial \ln \Gamma_{1}}{\partial \ln \rho}\right)_{Z,Y,P} \frac{\delta \rho}{\rho} + \left(\frac{\partial \ln \Gamma_{1}}{\partial Y}\right)_{Z,P,\rho}\delta Y \nonumber \\ & +\left(\frac{\partial \ln \Gamma_{1}}{\partial Z}\right)_{Y,P,\rho} \delta Z, \label{eqgammastate}
\end{align}
Two hypotheses are made when using helium kernels. First, one assumes that the equation of state of the reference model and that of the target model are the same. Secondly, one usually drops the last term in $\delta Z$ of Eq. (\ref{eqgammastate}). This is often considered to be a benign hypothesis and we will review its impact for various kernels in section \ref{SecNumerics}.
\\
\\
The problem of this method is that, when deriving the differential equation for the kernels, one may be faced with discontinuous terms within the equation coefficients. These discontinuities are due to the effects of the transition from radiative regions to convectives regions and have to be treated correctly if one does not wish to introduce numerical errors in the resolution. Typically, these discontinuities appear when taking first or second derivatives of the density (or any quantity related to the density through an algebraic relation). For example, the second derivative of the adiabatic squared sound speed, $c^{2}$ shows a discontinuity at the base of the convective envelope. This also means that the differential equation must be solved on separated domains and that continuity conditions have to be applied for each sub-domain. These conditions typically serve as constraints to solve the differential equations of structural kernels. For example, for the $(u,\Gamma_{1})$ and $(u,Y)$ kernels, the resolution of the second order differential equation uses one central boundary condition that is derived from the differential equation itself and one ``boundary'' condition that stems from the integral equation. Namely, one assumes that the kernels have to satisfy their integral equation at some point of the sub-domain. For the next sub-domain, a continuity relation on the kernels is derived, since they have to be continuous for continuous variables, and the integral relation is again used to obtain an additional condition for the sub-domain. 
\\
\\
In practice, the use of the integral relation for the additional condition is not trivial, since sometimes one can be confronted with integrals of the layers above the layer on which one wishes to solve the differential equation\footnote{Thus the arguments of the integral are unknown.}. The problem is even more complicated when facing separated domains. Thus, one has to find a workaround based on the linearity of the problem and ends up solving a system of two differential equations on each sub-domain, where the equations are simultaneously connected through continuity relations and integral equations. With a little algebra, this can be done using a simple direct solver and finite difference discretization (In our case, we used the prescriptions of \citet{ReeseFD} for the grid on which the equation is solved). This leads in practice to a good accuracy in the results when care is taken in the computation of the derivatives of the coefficients and of the already known kernels. Indeed, these derivatives can be a source of significant numerical noise when calculated on a reference model of poor quality or when the eigenfunctions have been computed with a poor accuracy. 
\subsection{Method of conjugated functions - Application to the $A$-$Y$ kernels}\label{SecMethodKosov}
The method of conjugated functions is quite different from what is done in the direct method, although the starting point is still the equality of two variational expressions for different structural variables. This method was presented for the first time in a paper by \citet{Elliott} in the context of helioseismology and a more thorough presentation of the method can be found in \citet{Kosovichev}. In this section, we recall the basis of the method and apply it to the derivation of new kernels associated with the $(A,Y)$ structural pair. 
\\
\\
The quantity $A$ is called the convective parameter and is closely related to the Brunt-Väisälä frequency. It is defined as follows:
\begin{align}
A=\frac{d \ln \rho}{d \ln r}-\frac{1}{\Gamma_{1}}\frac{d \ln P}{d \ln r}
\end{align}
This quantity has the convenient property to be zero in adiabatically stratified convective regions. It is also very sensitive to changes in depth of the base of the convective zone and changes in upper regions of convective envelopes. The problem we will define is thus related to determining the change of structural variables from $(\rho,Y)$ kernels to the $(A,Y)$ kernels. The $(\rho,Y)$ pair is a convenient starting point but one could choose other starting variables. Thus, our goal is to find the functions $K^{n,l}_{A,Y}$ and $K^{n,l}_{Y,A}$ for a given stellar model such that for two models that are sufficiently close to each other, we have:
\begin{align}
\frac{\delta \nu^{n,l}}{\nu^{n,l}}&=\int_{0}^{R}K^{n,l}_{\rho,Y}\frac{\delta \rho}{\rho} dr + \int_{0}^{R}K^{n,l}_{Y,\rho} \delta Y dr\nonumber \\
&=\int_{0}^{R}K^{n,l}_{A,Y}\delta A dr + \int_{0}^{R}K^{n,l}_{Y,A} \delta Y dr.
\end{align}
First, we have to relate the linear perturbation of $A$ to the other structural variables. In the approach of conjugated functions, one starts by defining a system of differential equations between the model perturbations, where one wishes to relate the different perturbed structural quantities found in the starting and final integral relations. In this particular case, one has a system of $3$ differential equations that relates all the quantities together. This system is written:
\begin{align}
r \frac{d}{dr}\left( \frac{\delta \rho}{\rho} \right) &= \delta A + \frac{G m}{r c^{2}} \frac{\partial \ln \Gamma_{1}}{\partial Y} \vert_{P,\rho} \delta Y + \frac{G m }{c^{2}r} \left[ \frac{\partial \ln \Gamma_{1}}{\partial \ln P} \vert_{\rho,Y}+1 \right] \frac{\delta P}{P}, \nonumber \\ 
&+ \frac{G m }{c^{2}r} \left[ \frac{\partial \ln \Gamma_{1}}{\partial \ln \rho} \vert_{P,Y}-1 \right] \frac{\delta \rho}{\rho} - \frac{G m}{r c^{2}}\frac{\delta m}{m}, \label{Eqdeltarho} \\
r \frac{d}{dr}\left( \frac{\delta m}{m} \right)&=\frac{4 \pi r^{3}\rho}{m} \left( \frac{\delta \rho}{\rho} - \frac{\delta m}{m}\right), \label{Eqdeltam}\\
r \frac{d}{dr} \left( \frac{\delta P}{P} \right)&=\frac{G m \rho}{r P} \left[ \frac{\delta P}{P} - \frac{\delta \rho}{\rho} - \frac{\delta m}{m} \right]. \label{EqdeltaS}
\end{align}
\\
\\
As we will show in this section, the method of conjugated functions uses equations closely related to the system presented above. A major advantage is that this approach leads to a system with simple coefficients, for which the problem of numerical derivatives will not be as important as for the direct method. However, this method uses more hypotheses than the direct method and is consequently less well-suited for asteroseismology. Typically, the problem stems from the boundary conditions that are used to close the system and select a unique solution. For the surface boundary conditions, we have to assume that the mass of the observed target is the same as the mass of the reference model. At first, we thought that only the mean density was required to be fitted to ensure a verification of the variational expression but we will see how we were mistaken in the next section. Indeed, it can be argued that kernels for structural pairs such as the $(A,\Gamma_{1})$ pair or the $(A,Y)$ pair will never offer a good accuracy in the asteroseismic case as will be illustrated in section \ref{SecNumerics}.

The goal of the method of conjugated functions is to determine the unknown tridimensional vector $\vec{v}=(v_{1},v_{2},v_{3})$, which is a conjugated function linked to the structural kernels (see Eq. \ref{EqKAY}), solution of the following system (related to the system of equations \ref{Eqdeltarho} to \ref{EqdeltaS}):
\begin{align}
-r \frac{d \vec{v}}{dr}-\vec{v} = \mathcal{A}^{T}\vec{v} + \mathcal{C}^{T}\vec{K}_{1}, \label{Eqvecv}
\end{align}
where we have used the following definitions:
\begin{align}
&\mathcal{A}= \left( \begin{matrix} \frac{Gm}{rc^{2}}\left[\frac{\partial \ln \Gamma_{1}}{\partial \ln \rho}\vert_{P,Y}-1 \right] & -\frac{Gm}{rc^{2}} & \frac{Gm}{r c^{2}} \left[\frac{\partial \ln \Gamma_{1}}{\partial \ln P}\vert_{\rho,Y} +1\right] \\
\frac{4 \pi r^{3} \rho}{m} & -\frac{4 \pi r^{3} \rho}{m} & 0 \\
\frac{-Gm \rho}{rP}& \frac{-G m \rho}{r P}& \frac{G m \rho}{r P}\end{matrix} \right),
\\
&\vec{K}_{1}=(K^{n,l}_{\rho,Y}, K^{n,l}_{Y, \rho}), \\
&\mathcal{C}=\left( \begin{matrix}
1 & 0 & 0 \\
0 & 0 & 0 
\end{matrix} \right).
\end{align}
We also introduce the following definitions:
\begin{align}
&\vec{x}= \left( \begin{matrix} 
\frac{\delta \rho}{\rho} \\
\frac{\delta m}{m} \\
\frac{\delta P}{P}
\end{matrix} \right),\; \;\;
\vec{s}_{1} =\left( \begin{matrix} 
\frac{\delta \rho}{\rho} \\
\delta Y
\end{matrix} \right), \;\;\;
\vec{s}_{2} =\left( \begin{matrix} 
\delta A \\
\delta Y
\end{matrix} \right), \\
&\mathcal{B}= \left( \begin{matrix} 1 & \frac{G m}{r c^{2}} \frac{\partial \ln \Gamma_{1}}{\partial Y}\vert_{P,\rho} \\
0 & 0 \\ 
0 & 0 \end{matrix} \right), \\
& \vec{K}_{2}=(K^{n,l}_{A,Y}, K^{n,l}_{Y, A}), \\
& \mathcal{D}=\left( \begin{matrix}
0 & 0  \\
0 & 1 
\end{matrix} \right). 
\end{align}
We use the following boundary conditions for $r=0$:
\begin{align}
3\tilde{v}_{1}(0) + 3 \tilde{v}_{2}(0) &= -\frac{K^{n,l}_{\rho,Y}}{r^{2}\rho}(0), \\
\tilde{v}_{3}(0)&=0,
\end{align}
with $\tilde{v}_{i}=\frac{v_{i}}{r^{2}\rho}$. Using $\tilde{v}_{i}$ as variables for the system is motivated by the central limit of the structural kernels, such as $K^{n,l}_{\rho,Y}$, which goes as $\mathcal{O}(r^{2})$ in central regions. These boundary conditions can be obtained from the limit as $r$ goes to $0$ of Eq. \ref{Eqvecv} itself similarly to what is presented in \citet{Unno} for the pulsation equations. The last boundary condition of equation \ref{Eqvecv}, at $r=R$ is defined as follows:
\begin{align}
\frac{\delta \rho}{\rho}(R)v_{1}(R)+\frac{\delta m}{m}(R)v_{2}(R)+\frac{\delta P}{P}(R)v_{3}(R)=0,
\end{align}
and results from the elimination of surface term in the integration by parts in Eq. \ref{EqvIntP} which can be changed using l'Hospital's theorem to avoid having to define $\frac{\delta P}{P}(R)$ with the equation of hydrostatic pressure, thus considering both $P(R)$ and $\delta P(R)$ to be $0$:
\begin{align}
\frac{\delta \rho}{\rho}(R)v_{1}(R)+\frac{\delta m}{m}(R)v_{2}(R)+\left(\frac{\delta \rho}{\rho}(R)+\frac{\delta m}{m}(R)\right)v_{3}(R)=0. \label{EqBoundSurf}
\end{align}
The main problem with this equation is that both the $\delta m$ and $\delta \rho$ terms are unknown, it is thus impossible to derive a simple boundary conditions and the components of \vec{v} without an additional hypothesis. In helioseismology, one states that the mass of the observed target is known and one ends up with a simple equation only with $\delta \rho$. One then simplifies the $\delta \rho$ term and ends up with the following simple relation:
\begin{align}
v_{1}(R)+v_{3}(R)=0.
\end{align}
This problem is intrinsic to the method of conjugated functions since one defines the boundary conditions of the system using an expression containing structural perturbations.

Using these definitions, it can be proven that if $\vec{v}$ is the solution to this problem, then the structural kernels $K^{n,l}_{A,Y}$ and $K^{n,l}_{Y,A}$, for each perturbation of the model defined by $\vec{x}$ and $\vec{s}_{2}$, can be determined using the following relations:
\begin{align}
K^{n,l}_{A,Y}&=v_{1}, \label{EqKAY}\\
K^{n,l}_{Y,A}&=K^{n,l}_{Y,\rho}+K^{n,l}_{A,Y}\frac{Gm}{rc^{2}}\frac{\partial \ln \Gamma_{1}}{\partial Y}\vert_{\rho, P}, \label{EqKYA}
\end{align}
To demonstrate this property, let us first rewrite the system of equations \ref{Eqdeltarho} to \ref{EqdeltaS} in its vector form using the definitions we have just introduced:
\begin{align}
r\frac{d\vec{x}}{dr} = \mathcal{A}\vec{x} + \mathcal{B} \vec{s}_{2} \label{EqStrucConj1},
\end{align}
One can also write a trivial matrix relation between vectors $\vec{s}_{1}$ and $\vec{s}_{2}$:
\begin{align}
\vec{s}_{1}=\mathcal{C}\vec{x}+\mathcal{D}\vec{s}_{2}\label{EqStrucConj2},
\end{align}
We now apply the scalar product of Eq. \ref{EqStrucConj1} and Eq. \ref{EqStrucConj2} with $\vec{v}$, defining the scalar product on the functional space as:
\begin{align}
<a,b> = \int_{0}^{R} a(r)b(r) dr,
\end{align}
which is done in this case for each component of $\vec{v}$ and $\vec{x}$. We then obtain:
\begin{align}
< \vec{v}, r \frac{d \vec{x}}{dr} > &= < \vec{v}, \mathcal{A} \vec{x} > + < \vec{v}, \mathcal{B} \vec{s}_{2} > , \\
 -< r \frac{d \vec{v}}{dr}+ \vec{v}, \vec{x} > + \left[r \vec{v}  . \vec{x}\right]_{0}^{R} &= < \vec{v}, \mathcal{A} \vec{x} > + < \vec{v}, \mathcal{B} \vec{s}_{2} >
\label{EqvIntP}
\end{align} 
where we have applied an integration by parts and thus obtained a differential equation for $\vec{v}$. If one considers both equations \ref{EqvIntP} and \ref{Eqvecv}, we obtain:
\begin{align}
< \vec{K}_{1}, \mathcal{C} \vec{x} > = < \vec{v}, \mathcal{B} \vec{s}_{2}>\label{EqCondv},
\end{align}
The new kernels can then be determined using:
\begin{align}
<\vec{K}_{2}, \vec{s}_{2}> &= <\vec{K}_{1},\mathcal{C}\vec{x}> + <\vec{K}_{1},\mathcal{D}\vec{s}_{2}> \\
& =  < \vec{v}, \mathcal{B} \vec{s}_{2}> + <\vec{K}_{1},\mathcal{D}\vec{s}_{2}>,
\end{align}
where we have used Eq. \ref{EqCondv}. If we develop the scalar products, we obtain the following integral relations:
\begin{align}
\int_{0}^{R} K^{n,l}_{A,Y}\delta A dr + \int_{0}^{R} K^{n,l}_{Y,A}\delta Y dr &= \int_{0}^{R} v_{1} (\delta A + \frac{G m}{r c^{2}} \frac{\partial \ln \Gamma_{1}}{\partial Y}\vert_{P,\rho}  \delta Y)dr\nonumber \\
& +\int_{0}^{R}K^{n,l}_{Y,\rho}\delta Y dr.
\end{align}
From these relations, we directly obtain the relations \ref{EqKAY} and \ref{EqKYA} and have thus demonstrated that determining the vector $\vec{v}$ satisfying Eq. \ref{Eqvecv} allowed us to determine the kernels of the $(A,Y)$ structural pair.

However, a few comments must be made on Eq. \ref{EqBoundSurf} since it leads to a strong limitation in the use of the method of conjugated functions. As previously stated, the boundary conditions applied are that the mass of the observed target and the reference model are the same. In asteroseismology, this is not necessarily the case. When using this method for other kernels, we could avoid this limitation by using the relation $\frac{\delta m}{m}=\frac{\delta \bar{\rho}}{\bar{\rho}}$ if the radius is fixed. Ultimately, one ends up with the same implicit scaling presented before for the direct method. It is a considerable advantage of the direct method that it does not explicitly uses any hypothesis on the mass of the observed target.

In this particular case, scaling the perturbations is impossible since the quantities $\delta A$ and $\delta Y$ are adimensional and not expressed as relative perturbations, obviously because $\frac{\delta A}{A}$ would be undetermined when $A$ goes to zero for the reference model. Consequently, the trick of the implicit scaling cannot be used and we are limited by the accuracy of radii determinations for asteroseismic targets.   

However, even with scaled models, the problem can still be present for the helium integrals. Indeed, for the density or the sound speed, the link is quickly done since these variables are explicitly part of what is called the acoustic structure of the stellar model and are directly linked to the oscillation frequencies. The question is more difficult when one thinks about the helium mass fraction. The problem is to link the helium mass fraction profile of the scaled target model to the helium profile of the real target. As such, there is no clear link between both profiles and helium cannot be directly related to the dynamical time since it is not an explicit variable of the acoustic structure. Therefore, caution as to be taken when determining helium abundances from inversion techniques when there is no strong constraints on the radius\footnote{As such the mass of the model would not be a problem if one considers that the mean density can be very accurately determined using seismology. Thus, if one knows the radius accurately, an accurate estimate of the mass can be determined provided good seismic data.}.

\begin{figure*}[t]
	\flushleft
		\includegraphics[width=18cm]{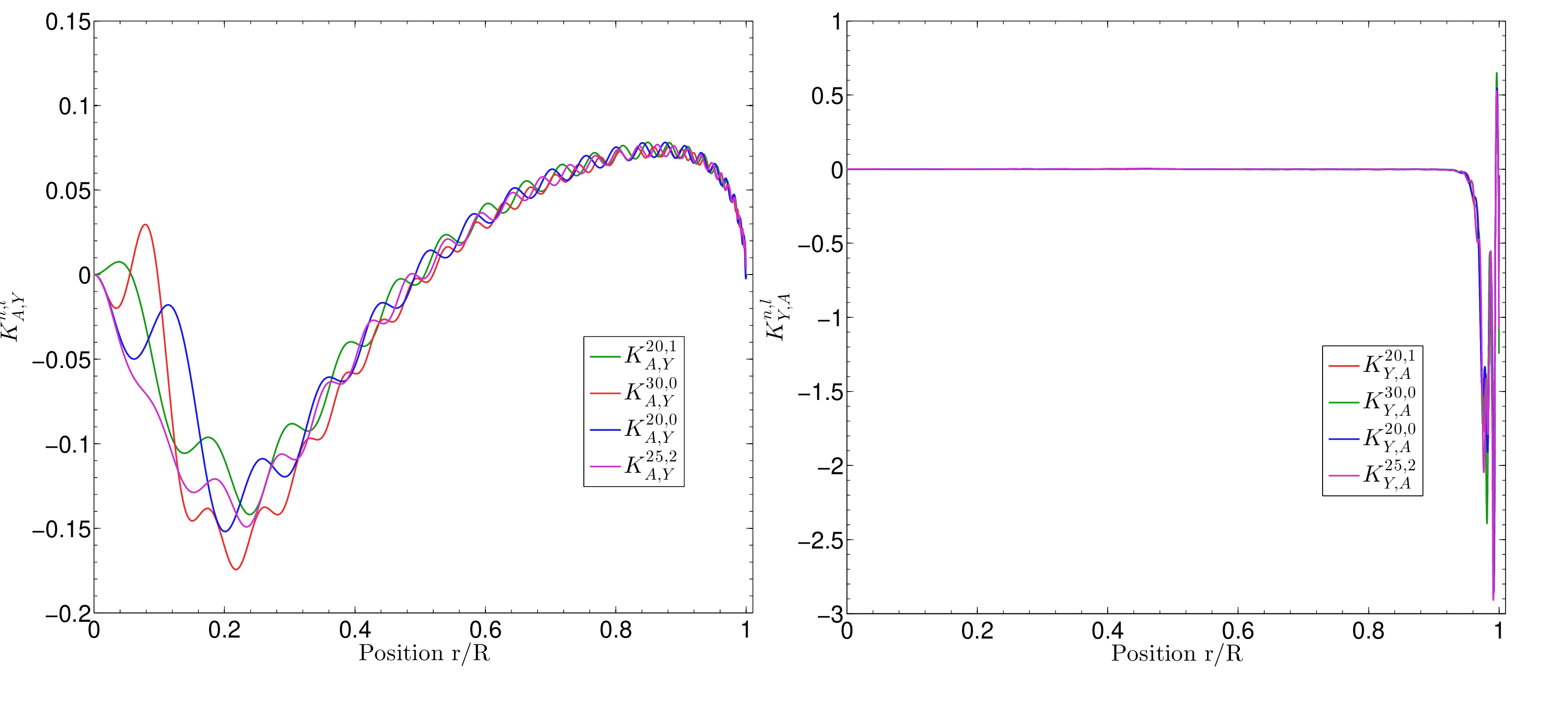}
	\caption{Illustration of various kernels for the $(A,Y)$ structural pairs for various degrees and radial orders.}
		\label{figKerAY}
\end{figure*}

We illustrate the $(A,Y)$ kernels in Fig. \ref{figKerAY} for various degrees and radial orders. It should be noted that the kernels associated with the convective parameter $A$ of the $(A,\Gamma_{1})$ structural pair are quite similar to the kernels associated with $A$ for the $(A,Y)$ structural pair and could thus be used to carry out inversions of similar indicators without the need to introduce the equation of state in the problem. The main problem is then to cope with the high amplitude of the cross-term kernels but ultimately, the presence of pairs of kernels with similar behaviours can be used to check the robustness of the inversion for observed data since it should lead to similar results if the cross-term is properly damped for both structural pairs. 
One additional striking feature of the $(A,Y)$ structural pair is the high amplitude of the helium kernels when compared to those of the convective parameter. It is pretty unusual since as was already noticed for the $(\rho,Y)$ structural pairs and confirmed for other structural kernels we derived, the helium kernels tend to have very low amplitudes and are thus very well adapted as cross-terms of inversions\footnote{Although errors on the equation of state can be non-negligible at the levels of accuracy of helioseismology.}. This makes these kernels very interesting for inversions of helium abundance using appropriated indicators in the solar case, where the data is abundant and the radius of the observed target is very well constrained and used to build the standard solar model used as a reference for the inversion. In terms of numerical quality, the verification of the initial system of differential equations is done up to relative differences of the order of $10^{-14}$ on the average. Typically the resolution is more accurate ($10^{-16}$ or less) in central regions and less accurate at the surface ($10^{-13}$). It should be noted that the numerical quality of the results is naturally still subject to the number of points of the models and the variables and unknowns considered in the system of equations.
\section{Numerical experiments}\label{SecNumerics}
In this section, we describe a few numerical experiments carried out to analyse the importance of various hypotheses used to compute structural kernels. All models were computed using the Clés stellar evolution code \citep{ScuflaireCles} with the following ingredients: the CEFF equation of state \citep{CEFF}, the OPAL opacities from \citet{OPAL}, supplemented at low temperature by the opacities of \citet{Ferguson} and the effects of conductivity from \citet{Potekhin} and \citet{Cassisi}. The nuclear reaction rates are those from the NACRE project \citep{Nacre}, supplemented by the updated reaction rate from \citet{Formicola} and convection was implemented using the classical, local mixing-length theory \citep{Bohm}. We also used the implementation of microscopic diffusion from \citet{Thoul}, for which three groups of elements are considered and treated separately: hydrogen, helium and the metals (all considered to have diffusion speeds of $^{56}Fe$). The oscillation frequencies and eigenfunctions were computed using the Liège adiabatic oscillation code \citep{ScuflaireLosc}. We took much care to analyse the numerical quality of the eigenfunctions and the models before computing structural kernels. Irregularities and poor quality of the computed eigenfunctions can bias the results and lead to wrong structural kernels and thus wrong inferences from inverted results. From our experience in hare-and-hounds exercises and inversions, we have determined that adding seismic constraints to the model is very efficient at bringing the reference model into the linear regime thus validating the inversion process. In other words, fitting the average large and small frequency separations is already a big improvement in terms of linearity, although individual seismic constraints, such as individual frequency ratios and individual small frequency separations are the best way to maximise the chances of being in the linear regime. Individual large frequency separations can also be used, but due to their sensitivity to surface effects, they should not be used in observed cases. As such, since in this study we did not use very elaborate seismic fitting techniques, our tests serve the only purpose of isolating various contributions to the errors and to test various hypotheses usually done when carrying out structural inversions in the context of helio- and asteroseismology.
\\
\\
We started by computing $4$ target models with different physical ingredients summarized in Table \ref{tabTarget}. Among these effects, we tested opacity changes, changes in the equation of state, the impact of the metallicity, the impact of individual abundance tables along with changes of typical parameters used for seismic fits such as the mixing-length parameter, $\alpha_{MLT}$ and the hydrogen abundance. For each target, we computed reference models with the same mass and similar physical ingredients. To ensure that both target models and reference models had the same radius, we used a minimization algorithm to fit the mean density of the target model varying the age of the reference model. In other words, since on the main-sequence the radius is changed due to slight core contraction and envelope expansion, we could ensure with this simple method a straightforward fit of all targets. Of course, this approach is limited. For instance, a model which includes efficient microscopic diffusion or a completely different chemical composition will not be strongly constrained by the fit of the mean density and thus will surely not be lying in the linear regime. This should be kept in mind throughout the paper since it is not what is done in typical seismic studies where all the available information is used. 
\begin{table*}[t]
\caption{Physical ingredients of the target models used for the hare-and-hounds exercises.}
\label{tabTarget}
  \centering
\begin{tabular}{r | c | c | c | c}
\hline \hline
 & \textbf{Target Model 1}& \textbf{Target Model 2}& \textbf{Target Model 3}& \textbf{Target Model 4} \\ \hline
\textit{Mass ($\mathrm{M_{\odot}}$)}& $1.0$ & $1.0$ &$1.0$&$1.0$\\
\textit{Radius ($\mathrm{R_{\odot}}$)}& $1.0712$  &$1.0822$& $1.0394$&$1.0770$\\ 
\textit{Age ($\mathrm{Gyr}$)} &$5.0$ & $5.0$ &$5.0$&$4.5$\\ 
\textit{EOS} & CEFF & OPAL & CEFF&OPAL\\
\textit{Abundances} & GN$93$ & GN$93$& AGSS$09$&AGSS$09$\\
\textit{$X_{0}$} &$0.7$&$0.7$&$0.7$&$0.67$\\
\textit{$X_{0}$} &$0.015$&$0.015$&$0.015$&$0.02$\\
\textit{$\alpha_{\mathrm{MLT}}$} & $1.7$ &$1.7$&$1.7$&$1.7$\\
\textit{Mixing} & $-$ &$-$&$-$&Settling+turbulent diffusion\\
\hline
\end{tabular}
\end{table*}
The verification of the linear relations between frequencies and structural profiles is demonstrated by plotting the relative differences between the right-hand side and left-hand side of the linear integral relations, denoted $\mathcal{E}^{n,l}_{s_{1},s_{2}}$ defined as follows:
\begin{align}
\mathcal{E}^{n,l}_{s_{1},s_{2}}=\frac{\frac{\delta \nu^{n,l}}{\nu^{n,l}}-\left( \int_{0}^{R} K^{n,l}_{s_{1},s_{2}}\frac{\delta s_{1}}{s_{1}}dr + \int_{0}^{R} K^{n,l}_{s_{2},s_{1}}\frac{\delta s_{2}}{s_{2}}dr \right)}{\frac{\delta \nu^{n,l}}{\nu^{n,l}}}
\end{align}
with $s_{1}$ and $s_{2}$ being any of the structural variables for which structural kernels can be obtained. Using this approach offers a straightforward method to compare the validity of the linear relations for each kernel and each mode, pointing out possible weaknesses and inaccuracies. For each comparison, we used the modes with $\ell=0$, $1$, $2$ and $3$ and $n$ between $6$ and $41$. 
\subsection{Limits of the linear regime}
\begin{figure*}[t]
	\flushleft
		\includegraphics[width=18cm]{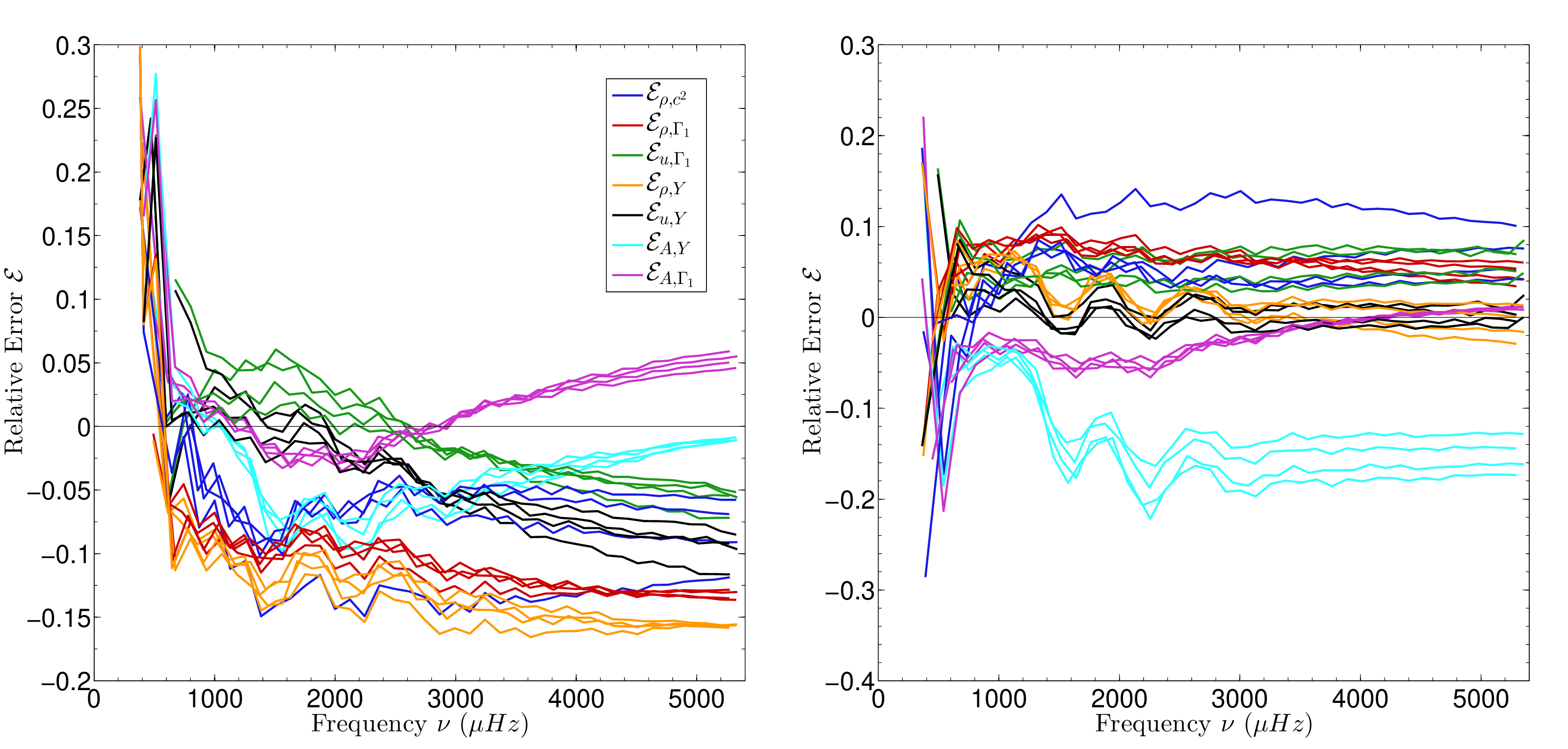}
	\caption{Left panel: verification of the linear integral relations between structure and frequencies for target $1$ and a model with $\alpha_{MLT}=1.5$ and $X_{0}=0.69$. Right panel: same as left panel but for a model with $\alpha_{MLT}=1.9$ and $X_{0}=0.715$. We clearly see that different structural pairs do not satisfy the linear relations to within the same accuracy. Each curve corresponds to a given $\ell$ of the modes.}
		\label{figVar}
\end{figure*}

First of all, we illustrate in Figure \ref{figVar} the verification of the linear relations between frequency differences and structural differences for various structural pairs for target $1$ and two reference models, with slightly different $\alpha_{MLT}$ and $X_{0}$ values. The model associated with the left panel has $\alpha_{MLT}=1.5$ and $X_{0}=0.69$, whereas the model used as a reference for the right panel has $\alpha_{MLT}=1.9$ and $X_{0}=0.715$. We can also see that all structural pairs do not satisfy the linear relations to within the same accuracy. Typically, kernels for the convective parameter $A$ can be problematic, especially kernels of the $(A,Y)$ structural pair. This can mean that all perturbations of the quantities may not be in the linear regime, and that for some kernels, second order terms should be considered. Ultimately this can be the case for variables other than the convective parameter and the $(A,Y)$ kernels can sometimes satisfy the linear relations whilst the $(\rho,c^{2})$ kernels do not. Two other problems of hare-and-hounds exercises using various kernels have to be mentioned: first, the insufficient numerical quality of the model and of the eigenfunction themselves; second, the changes of the parameters of the models can sometimes be inappropriate to test these relations and thus, inversion techniques. This means that we are intrinsically limited in our tests for robustness of inversions and that to some extent, other approaches could be sought to fully constrain the limitations of inversions in the context of asteroseismology.
The first point is quite straightforward and linked to various problems that can be found in stellar evolution codes. For example, the quality of numerical derivatives, which is a function of both the derivation scheme that is used and the quality of the grid on which the model or the eigenfunction is computed. Another highly underestimated error is the finite accuracy with which a stellar evolutionary model satisfies hydrostatic equilibrium. In other words, the intrinsic consistency of thermodynamical quantities used to describe the acoustic structure of the model must be checked. To these two sources of errors, we must add the possible differences stemming from intrinsic methods used to compute the models in various stellar evolutionary codes.

\begin{figure*}[t]
	\flushleft
		\includegraphics[width=18cm]{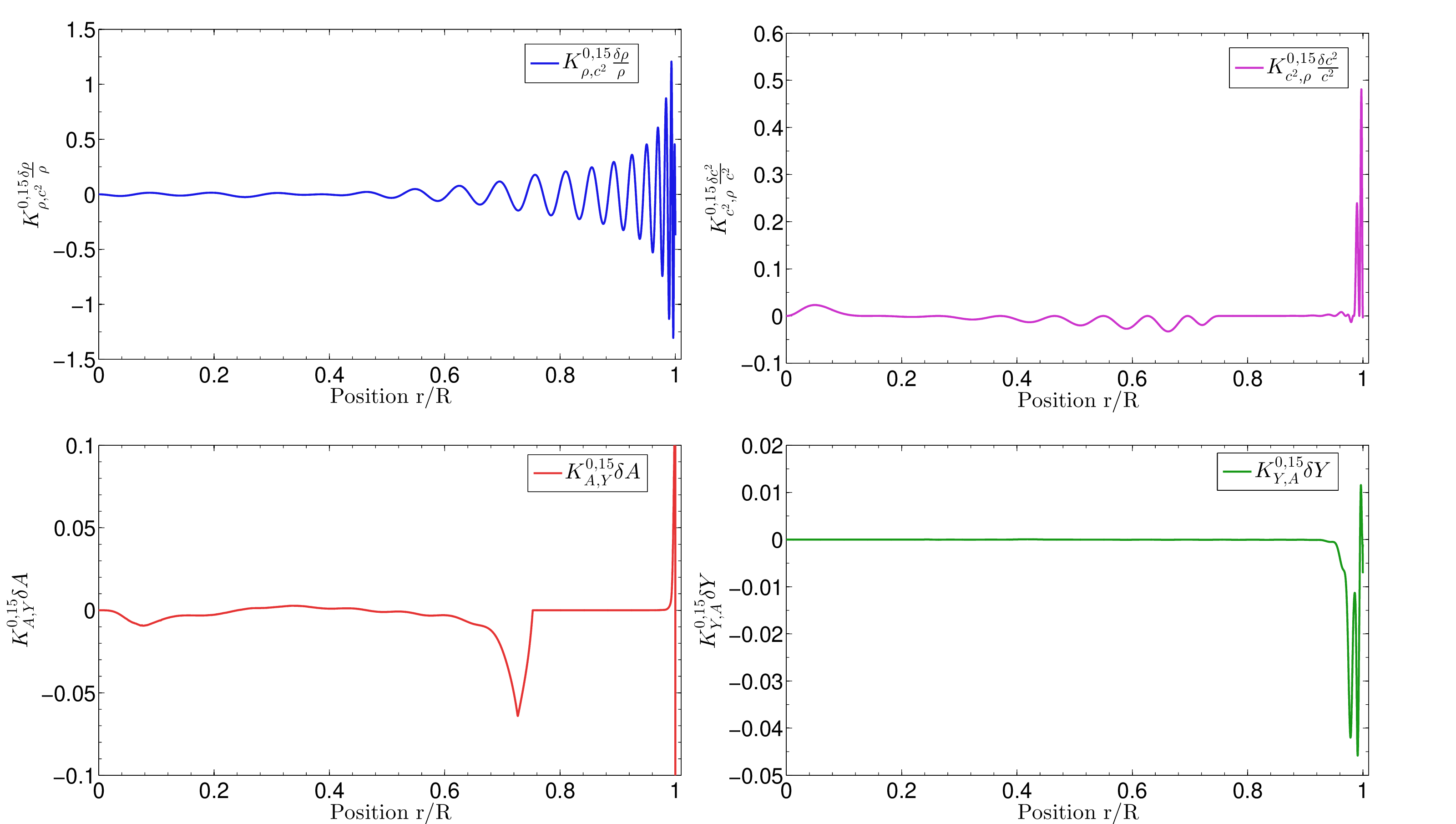}
	\caption{Arguments of the integrals of the linear relations between frequency and structure. $\rho$ argument (upper left) and $c^{2}$ argument (upper right) from the $(\rho,c^{2})$ pair. $A$ argument (lower left) and $Y$ argument (lower right) from the $(A,Y)$ pair. Each curve corresponds to a given $\ell$ of the modes.}
		\label{figProfStruc}
\end{figure*}

Intrinsic non-linearity is a recurring problem when using the frequency structure relations. In figure \ref{figProfStruc}, we illustrate the arguments of the structural integrals from the $(\rho,c^{2})$ pair and  $(A,Y)$ pair. The $\rho$ and $c^{2}$ arguments have very regular patterns naturally more concentrated towards the surface regions due to the higher amplitude of the kernels. Similarly, the amplitude of the $Y$ contribution in the lower right panel is only important in the surface regions. Although smaller than the other contributions, this helium integral is a factor $2$ larger than the helium integral from the $(\rho,Y)$ structural pair. As we will see later, this has important implications for the limitations of the linear regime with the $(A,Y)$ kernels. In the lower-left panel of Fig. \ref{figProfStruc}, we can see that the $A$ term is much more important in the surface regions, with a small contribution coming from the base of the convective envelope. This means that in practice, this structural pair might well be very sensitive to surface effects. From the numerical point of view, this means that to use the $(A,Y)$ or $(A,\Gamma_{1})$ pair, a very good quality of the grid as well as of the structure equations in the uppermost regions of the model is necessary to avoid important numerical uncertainties. We emphasize here that being able to build the $(\rho,c^{2})$ pair to within a good accuracy does not mean that numerical errors remain small when building new kernels from the existing ones. This is particularly true for the $(A,Y)$ and $(A,\Gamma_{1})$ kernels but can also be seen for other pairs.

Another extreme is the case where the perturbation of certain thermodynamic quantities can be considered small and thus within the linear regime while other cannot. In this case, certain linear relations might be valid while others are not. The case can be illustrated with kernels related to helium. Let us take two models, with the same mass, radius, chemical composition and mixing-length parameter. In one of the models, we include microscopic diffusion but not with its full intensity by multiplying the diffusion speeds by a factor $D$ smaller than one (here for example, we chose $0.5$). The surface helium abundance has significantly changed. We see a difference in mass fraction of the order of $0.025$, in other words, nearly $10 \%$. It is obvious that the changes cannot be considered small and it is then no surprise to see that these models are within the linear regime for the $(A,\Gamma_{1})$ kernels but not the $(A,Y)$ kernels. Again this means that caution is required when changing the structural pair in an inversion process and that usually, the validity of the linear regime can be assessed by using different reference models to carry out the inversion with one structural pair. Ultimately, if the inversion result, let us say, changes significantly with the structural pair that is used, then there is a problem with the inversion process. In the case of the $Y$ kernels, the problem can also arise due to the assumption that the equation of state is known, since it is used to derive the kernels. In these test cases, we always used the same equation of state for both target and reference model, except when it is specifically mentioned as in the other test cases below.

However, even when the equation of state is the same, we noticed that Eq. \ref{Eqdeltarho} is not always perfectly satisfied. If the same equation is written for the $(A,\Gamma_{1})$ kernels, then the agreement is improved, meaning that some of the errors seen for the $(A,Y)$ kernels can be attributed to the verification of Eq. \ref{eqgammastate}. This hypothesis has been tested and we clearly saw a disagreement between the left-hand side and the right-hand side of equation \ref{eqgammastate}. This disagreement did not seem to arise from numerical uncertainties but rather from the intrinsic non-linearity of the equation, due for example to shifts in the ionisation zones that were not reproduced by the linear expansion with derivatives of $\Gamma_{1}$. Knowing this, it thus seems perfectly normal to see a stronger non-linear behaviour for the $(A,Y)$ kernels since they have a much higher helium contribution than other kernels. This of course implies limitations on direct helium determinations from kernel inversions and requires further investigation. 
\subsection{Effects of metallicity}
\begin{figure*}[t]
	\flushleft
		\includegraphics[width=18cm]{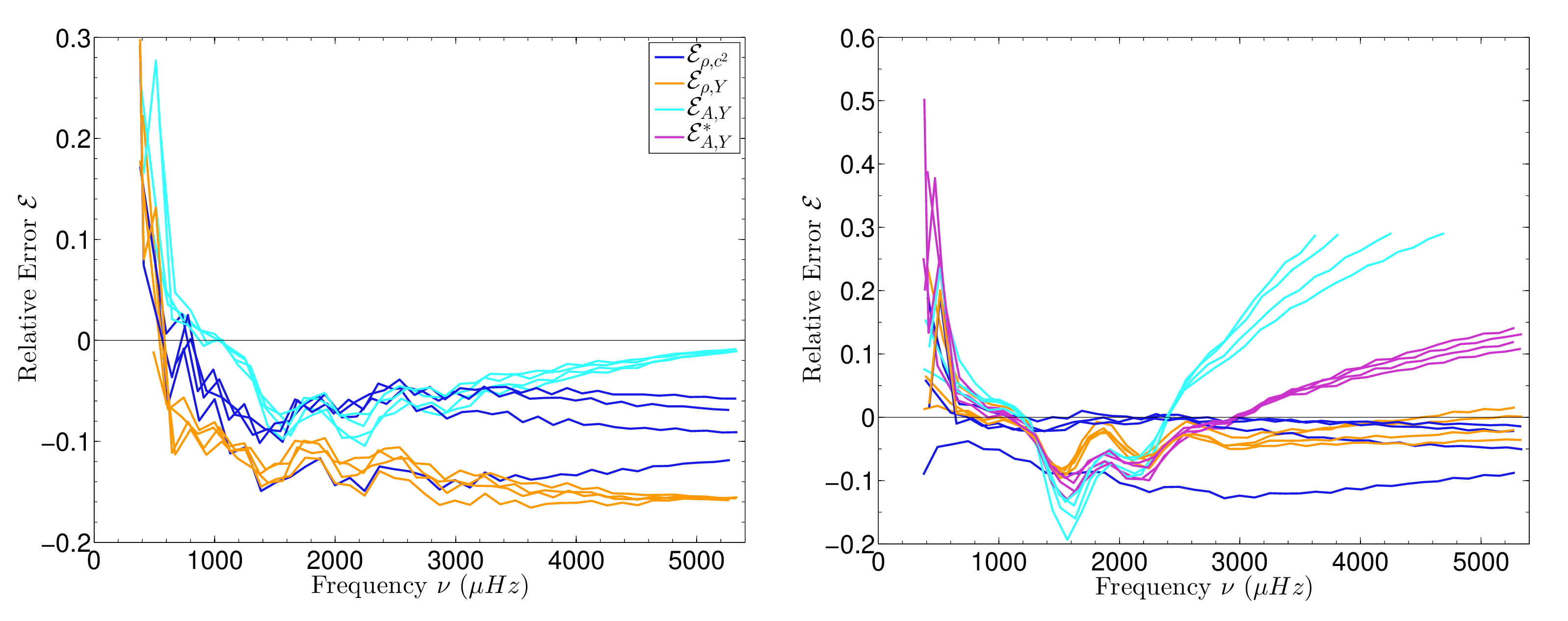}
	\caption{Left panel: Verification of the linear integral relation for a reference model which has the same metallicity as target $1$. Right panel: effects of a $0.002$ shift to the metallicity on the verification of the linear integral relations. The effects of a $0.001$ shift to the metallicity for the $(A,Y)$ kernels only is shown in purple and referenced with a $*$. Each curve corresponds to a given $\ell$ of the modes.}
		\label{figVarZ}
\end{figure*}

The effect of metallicity are extremely important to quantify since they are often neglected when trying to assess the helium abundance. Using a few models, we review the impact of small changes of metallicity on the verification of the linear relations. This impact is illustrated in figure \ref{figVarZ}. An important point to mention is that in asteroseismic observed cases, the metallicity is calculated through the spectroscopic observations of $\left[ Fe / H \right]$. One then uses the sun as a reference but it should be emphasized that there is no agreement to this day on the solar metallicity and that this uncertainty as such has an impact on linear relations, especially when using $Y$ related kernels.

In figure \ref{figVarZ}, we can disentangle the impact of metallicity, since on the left-hand plot, both the target and reference models have the same $Z$, whereas on the right-hand plot, we changed the metallicity by $0.002$. Of course, since the models do not have the same age, some changes can be seen due to intrinsic differences in the models, but it is still striking to see that the difference in $Z$ can have an impact in some contexts, which is in contradiction with what was previously believed. The case of the high frequency range of the $(A,Y)$ kernel is a very good illustration of how this can be a problem. However, we note that other kernels, such as the $(A,\Gamma_{1})$ pair were affected by the changes in metallicity, but not as much, so the intrinsic differences coming from $Z$ is at least a few per cent. From inspection of the behaviour of the $(\rho, \Gamma_{1})$ pair, we can say that the differences for the $(\rho,Y)$ pair also stem from intrinsic differences and not only from the term in $\delta Z$ neglected in Eq. \ref{eqgammastate}. This means that metallicity can be extremely important for some fitting processes in terms of the  validity of the linear structural relations due to the intrinsic differences that can be generated between the target and reference models.

We also plotted in purple the verification of the linear relations for a model with a metallicity change of $0.001$. We can see that the errors are divided by approximatively a factor $2$. This means that the effect of the metallicity is rather global and goes beyond the neglect of the additional term in Eq. \ref{eqgammastate}. This is further confirmed by the impact of the metallicity on the $(A,\Gamma_{1})$ structural pair, where the $\delta Z$ contribution is not explicitely involved. Figure \ref{figVarZ} also shows that the $(\rho,c^{2})$ kernels are not affected by metallicity changes, as expected, thus leaving their diagnostic potential unaltered.
\subsection{Effects of the equation of state}
\begin{figure*}[t]
	\flushleft
		\includegraphics[width=18cm]{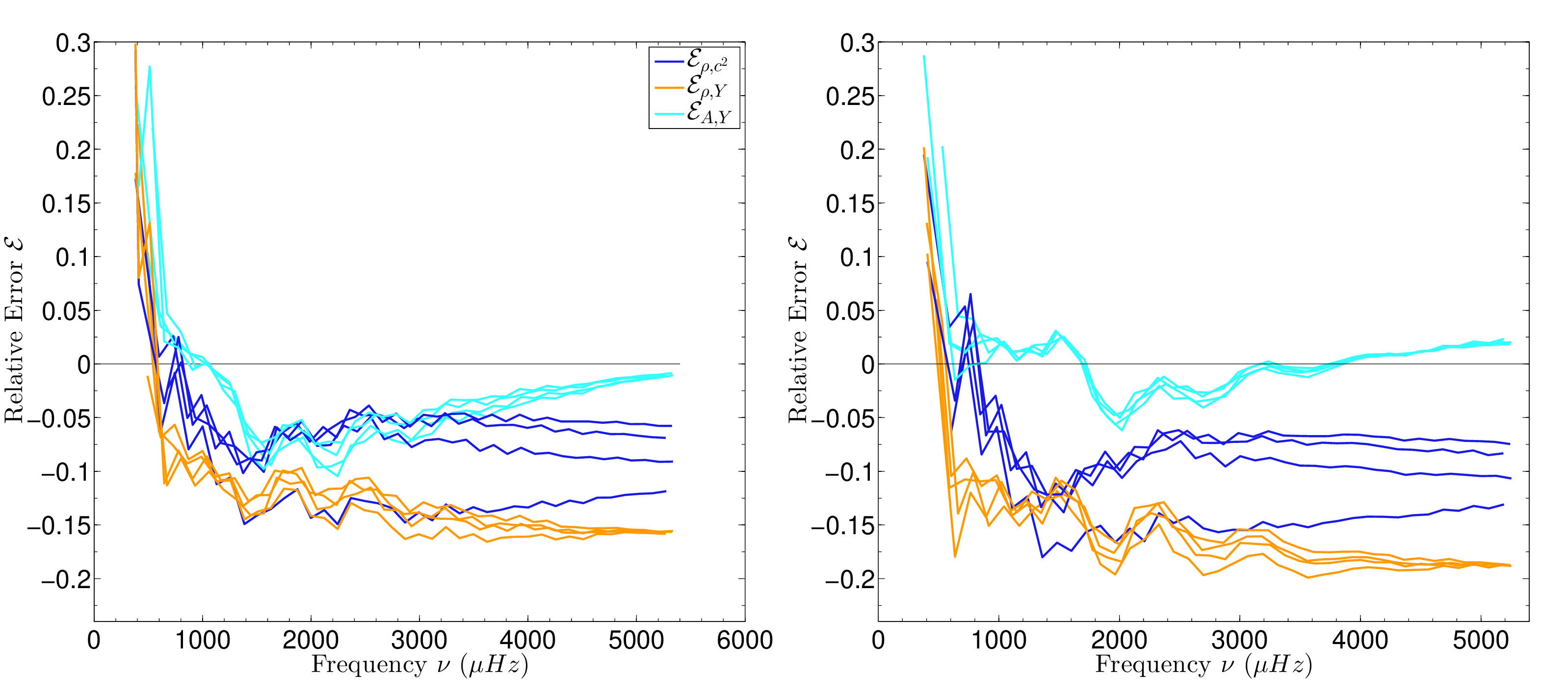}
	\caption{Left panel: Verification of the linear integral relation for a reference model which has the same equation of state as target $1$. Right panel: effects of a change from the CEFF equation of state on the OPAL equation of state to the verification of the linear integral relations. Each curve corresponds to a given $\ell$ of the modes.}
		\label{figVarEOS}
\end{figure*}

In figure \ref{figVarEOS}, we illustrate the same plot as figure \ref{figVarZ}, but changing the equation of state of the target model to the Opal equation of state \citep{OPALEOS}. The reference models are built with the CEFF equation of state which is used to compute the derivatives of $\Gamma_{1}$ and derive consistent variational expressions. Figure \ref{figVarEOS} thus illustrates the impact that not knowing the equation of state of the target has on the verification of the linear structural relations. While there is some impact, it is not as large as expected. 
\\
\\
Of course this does not mean in any case that the equation of state is not important for the linear integral relations used in inversions, but when compared to the impact of metallicity, it seems that in this case $Z$ has a larger impact. This is not to be generalized but means that we have to be careful with the approximations made and perhaps, in the case of the solar metallicity problem, both the uncertainties on metallicity and the equation of state have to be taken into account. One point worth mentionning about this test case is that both the CEFF and the OPAL equations of state are very similar for solar conditions, so the small impact is a result of similarities between theoretical equations of state and might not be representative of the differences between the true equation of state in the sun and one of the theoretical ones.
\subsection{Effects of abundances and radii inaccuracies}
\begin{figure*}[t]
	\flushleft
		\includegraphics[width=18cm]{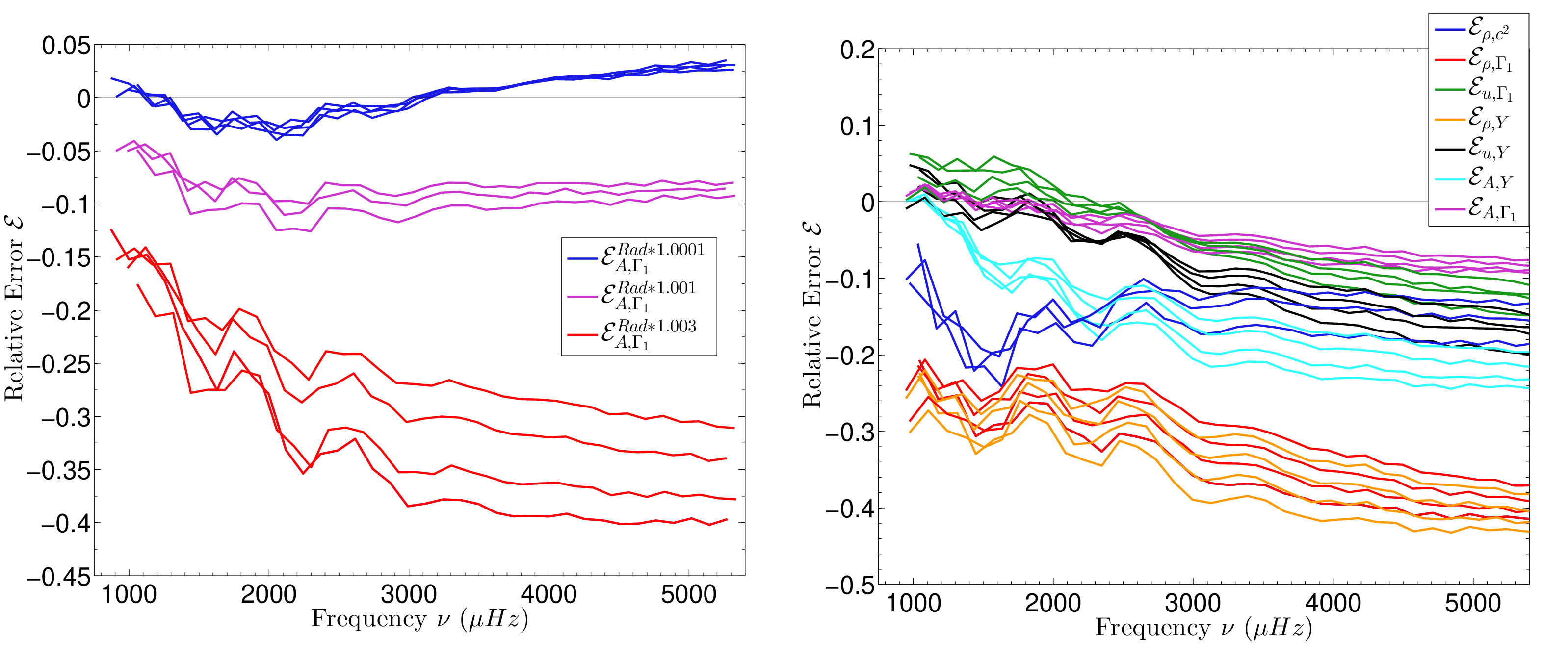}
	\caption{Left panel: effect of mismatches in radii between the reference and target models. The relative values of the mismatches are respectively $10^{-4}$ for the blue dots, $10^{3}$ for the magenta dots and $3\times 10^{-3}$ for the red dots. Right panel: effects of changes in the abundances of heavy elements for different structural pairs. Each curve corresponds to a given $\ell$ of the modes.}
		\label{figVarRad}
\end{figure*}

To test the impact of microphysics, we computed the target model with the AGSS$09$ \citep{Asplund} heavy elements mixture and the same metallicity as the reference model, computed with the GN$93$ abundances \citep{Grevesse}. Indeed, this changes significantly the thermodynamic quantities inside the star and affects significantly the opacity. As such, this test can be seen as one way to demonstrate that microphysics also has a large impact on the verification of the linear integral relation used to carry out inversions. From the right hand side panel of Fig. \ref{figVarRad}, we can see that all the kernels are affected by the microphysics. The effects are mostly seen for the $(\rho,c^{2})$, $(\rho,\Gamma_{1})$ and $(\rho,Y)$ pairs, but the good results of the other pairs are likely due to chance since all profiles have been significantly affected by the modified microphysics. 
The problem of the radius fit that was discussed in section \ref{SecMethodKosov} is illustrated on the left side panel of figure \ref{figVarRad}. The tendency is clearly seen since introducing progressively an error on the mean density produces an important error on the verification of the linear relations. The problem would be similar if one would consider the mean density to be known within an excellent accuracy but the mass to be unknown. In such case, a small error on the mass introduces an error of the order of $R^{3}$ thus an even larger departure from the linear integral relation. The problem is intrinsically due to the adimensional nature of $A$, meaning that it cannot be scaled to take into account our ignorance of the mass or radius of the target. Indeed, this effect is not seen for any kernel computed with the direct method if the proper scaling is applied to the structural variables. This leads to intrinsic limitations of the application of the $(A,\Gamma_{1})$ and $(A,Y)$ kernels in asteroseismology. The problem may not be solved in this case by changing the fitting method since fitting seismic constraints may not always ensure a good fit of the radius of the observed target. Ultimately, these kernels are only limited to the very best asteroseismic targets for which excellent interferometric measures of the radii are available. 
\subsection{Effects of extra-mixing}
\begin{figure*}[t]
	\flushleft
		\includegraphics[width=18cm]{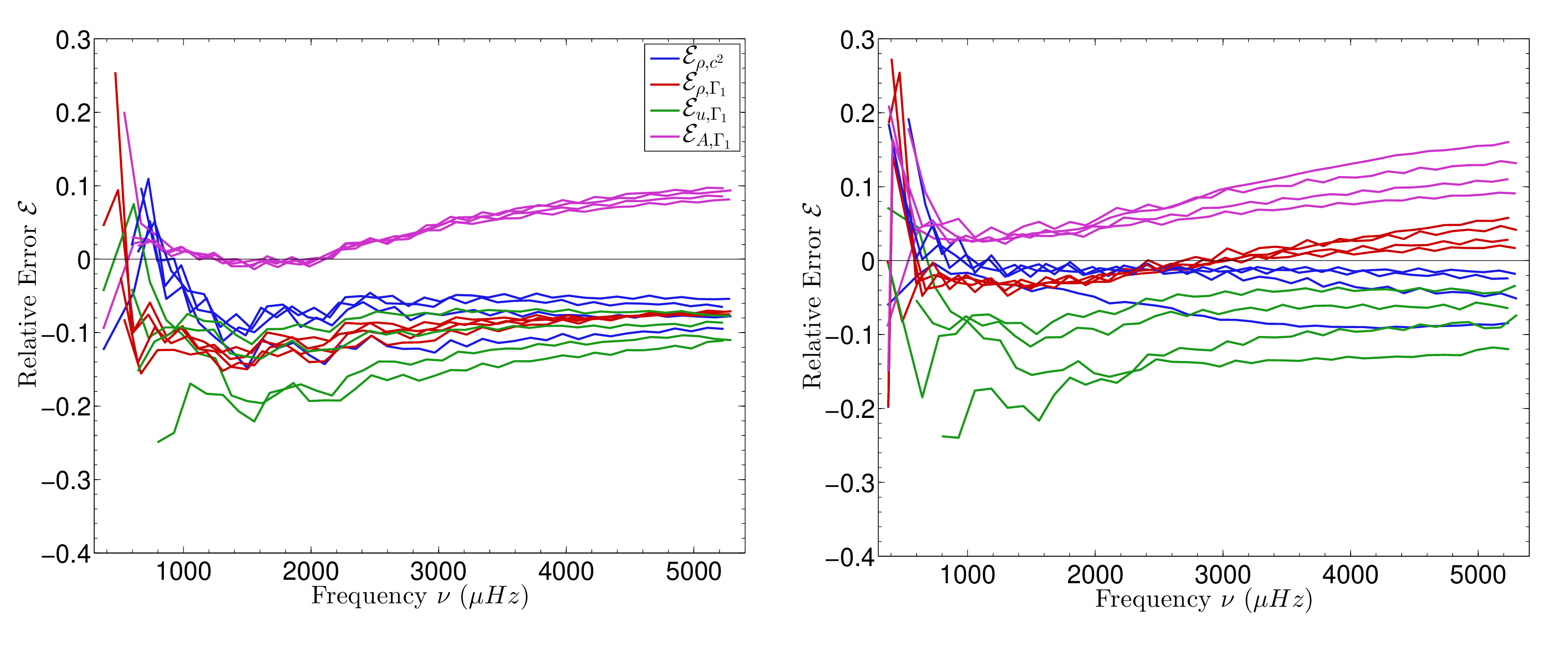}
	\caption{Effect of extra mixing on the verification of linear structural relations for both models. The left panel is associated with a model with a slightly inhibited extra-mixing intensity, whereas the right panel has a nearly constant extra-mixing throughout the model layers, but with an even smaller intensity. Each curve corresponds to a given $\ell$ of the modes.}
		\label{figVarTurb}
\end{figure*}

The extra-mixing term is very common in stellar physics. It is used to introduce additional hydrodynamical processes not taken into account in standard stellar models. The problem of lithium abundances is a good illustration that some extra-mixing is actually taking place in real stars. Thus, it seems perfectly normal to ask the question whether the non-inclusion of additional mixing processes could affect the verification of linear integral relation between frequencies and structural quantities. The answer to this question is illustrated in Fig. \ref{figVarTurb}. To carry out this test case, we used target model $4$ which includes turbulent diffusion in addition to microscopic diffusion. To test the robustness of the linear relations, we used reference models for which turbulent diffusion had been inhibited. For example, in the left panel of Fig. \ref{figVarTurb}, the reference model had a slightly higher helium of $0.005$ abundance and a less efficient turbulent diffusion. The verification of the linear relations is still good, but it seems that the $(A,\Gamma_{1})$ and the $(\rho,\Gamma_{1})$ structural pairs are strongly affected by the neglect of extra-mixing. This statement is confirmed when looking at the right panel of Fig. \ref{figVarTurb} for which the reference model has a nearly constant extra-mixing throughout all layers of the model. The $(\rho,c^{2})$  and $(u,\Gamma_{1})$ kernels seem not to be too much affected by extra-mixing. In general, the impact of extra-mixing is much reduced for the models we tested here. This statement, of course, only applies for physical conditions similar to solar and for the fitting process we use in these numerical tests. This does not mean that the problem could not reappear for models with convective cores, for which extra mixing could change significantly the evolutionary path and the acoustic structure. 
\section{Conclusion}
In this paper, we analysed the verification of the linear integral relations between frequency and structural quantities frequently used in helioseismology for various structural pairs. In Sect. \ref{SecMethodDirect} and Sect. \ref{SecMethodKosov}, we presented the two methods to change the structural variables in the linear integral relations. The direct method we present in this paper has the advantage of being more general than the method of conjugated functions, which explicitly uses the same radius for the observed target and the reference model. Although this is implied in the direct method, it is not used as a pre-requisite in the derivation of the equations leading to the kernels. Moreover, it has been shown that it is possible circumvent the problem by re-scaling the information provided by the inversion \citep[see][for a discussion of this problem]{Buldgentu,BasuSca}. In that sense, the method we propose offers a good alternative that is applicable in the asteroseismic case. However, we have also shown that the method we propose leads to somewhat complicated coefficients which can be difficult to derive if the numerical quality of the model is not ensured. 
\\
\\
Furthermore, in Sect. \ref{SecMethodKosov}, we showed how the conjugated functions approach could be used to derive $(A,Y)$ kernels. These kernels have the particularity of showing a very high sensitivity to the chemical composition since they are the only structural kernels for which the helium kernels have a higher amplitude than those associated with the secondary variable of the structural pair. This property is extremely important in the context of structural inversions since the amplitude of helium kernels was the main motivation behind their use as a cross-term in helioseismology but also their main handicap for direct kernel-based inferences of the helium profile using classical inversion techniques such as the SOLA or the RLS method. However, by no means would these inversions be independent of the equation of state since it is introduced in the very equations leading to the $(A,Y)$ kernels.
\\
\\
In Sect. \ref{SecNumerics}, we presented various experiments showing the intrinsic limitations in the linear regime of structural pairs. These limitations can be due to numerical inaccuracies or to the intrinsic non-linear behaviour of different variables. The most striking example is that of helium, for which extra-mixing can change significantly the local abundance while hardly changing the sound speed or density profile. In that sense, the numerical experiments we presented, although intrinsically limited, show that changing the structural pair in the integral relations is not often innocent, especially at the verge of non-linearity\footnote{Which may well be the case in the context of asteroseismic inversions.}. 
\\
\\
In addition, we analysed the importance of various structural changes, such as the impact of metallicity changes. We showed that small changes of metallicity could affect significantly the linear structural relations, especially for the $(A,Y)$ structural pair. However, we stress here that the linear behaviour of the integral relations is strongly dependent on the fitting process. This emphasizes again that in the asteroseismic case, all information available should be used to ensure the verification of the linear structural relations. 
\\
\\
In addition to the effect of metallicity, we also analysed the impact of the equation of state on the integral relations. Surprisingly, we find them to be less important than previously stated and even less important in some cases than the metallicity effects on the verification of the linear relations. This could have an impact on the potential of inversion techniques in the solar case. Changing the equation of state has little impact, and may be the result of a bias from the fitting process we used or it may be due to the similarity under solar conditions of most equations of state. 
\\
\\
We also analysed the importance of radius constraints and constraints on the microphysics by changing the heavy elements mixture. The test case on the radius inaccuracies shows the importance of this additional constraint for kernels derived with the method of conjugated functions, whereas the kernels derived with the direct method are found to be more robust if the proper scaling is applied when analysing the inversion results. However, we also emphasize that adimensional variables, like $A$ or $Y$, cannot be rescaled. The test case on the heavy elements mixture showed the important sensitivity to microphysics in stellar models. As such, the reference and target models were quite different and it is not surprising that linear structural relations are strongly affected. However, from our experience in seismic modelling, we know that these differences can be reduced by introducing additional constraints. Ultimately, models with different abundances can be very similar in terms of thermodynamical quantities due to compensations. 
\\
\\
Finally, we also analysed the importance of additional mixing acting during the evolution of the target model. We found that for solar conditions, additional extra-mixing processes could change slightly the verification of the linear relations, but that these changes were not as significant as those obtained from inaccuracies in metallicity, for example. We stress that this analysis should be extended to other parts of the HR diagram, where extra-mixing can have a more significant impact on the acoustic structure of stellar models, and thus on the verification of the linear relations between frequencies and structure. 
\\
\\
To conclude, the advent of the space photometry era and the quality of data provided by past and upcoming space missions will allow us to use new seismic approaches to extract efficiently seismic information. However, it is still important to provide a theoretical framework for these methods, to test their limitations and to determine what additional information (spectroscopic, interferometric, ...) or methodological improvement are necessary to enable the use of seismic inversions in asteroseismology. As such, this study only gives answers to limited theoretical questions and is only one step towards the improvement of our use of seismic information. 
\begin{acknowledgements}
G.B. is supported by the FNRS (``Fonds National de la Recherche Scientifique'') through a FRIA (``Fonds pour la Formation à la Recherche dans l'Industrie et l'Agriculture'') doctoral fellowship. This article made use of an adapted version of InversionKit, a software developed in the context of the HELAS and SPACEINN networks, funded by the European Commissions's Sixth and Seventh Framework Programmes.
\end{acknowledgements}
\bibliography{biblioarticle5}
\appendix
\section{Convective parameter kernels from the direct method}\label{SecAYDirect}
As stated in the core of the paper, kernels for the structural pair $(A,Y)$ or $(A,\Gamma_{1})$ can be also obtained from the direct method. We give a few steps in the derivation of the third order differential equation that leads to these kernels and discuss a few problems regarding its numerical resolution. The first step is to introduce the helium and $A$ perturbations in the linear integral equation.
\begin{align}
\int_{0}^{R} K^{n,l}_{\rho,Y}\frac{\delta \rho}{\rho} dr + \int_{0}^{R}& K^{n,l}_{Y,\rho}\delta Y dr = \int_{0}^{R}K^{n,l}_{A,Y}\delta A dr + \int_{0}^{R}K^{n,l}_{Y,A}\delta Y dr \nonumber \\
=& \int_{0}^{R} K^{n,l}_{A,Y}\left[ r\frac{d}{dr}\left( \frac{\delta \rho}{\rho} \right)+\frac{Gm}{rc^{2}}\frac{\delta m}{m}\right. \nonumber \\
& -\frac{Gm}{rc^{2}}\left(\frac{\partial \ln \Gamma_{1}}{\partial \ln \rho}\vert_{P,Y} -1 \right) \frac{\delta \rho}{\rho} \nonumber \\
&\left. -\frac{Gm}{rc^{2}}\left(\frac{\partial \ln \Gamma_{1}}{\partial \ln P}\vert_{\rho,Y} +1 \right)\frac{\delta P}{P}\right]dr \nonumber \\
&+ \int_{0}^{R}\left[K^{n,l}_{Y,A}- K^{n,l}_{A,Y} \frac{Gm}{r c^{2}} \frac{\partial \ln \Gamma_{1}}{\partial Y}\vert_{\rho,P}\right]\delta Y dr.
\end{align}
Then, we have to use the definition of hydrostatic pressure and mass and permute the integrals. We can already notice that the term with the derivative of density will be problematic and will require an integration by parts. 
\begin{align}
&\left[ K^{n,l}_{A,Y} r\frac{\delta \rho}{\rho}\right]^{R}_{0} - \int_{0}^{R}\frac{d(rK^{n,l}_{A,Y})}{dr} \frac{\delta \rho}{\rho}dr \nonumber \\
& +\int_{0}^{R} 4 \pi r^{2} \rho \left[ \int_{r}^{R} \frac{GK^{n,l}_{A,Y}}{\bar{r}c^{2}}d\bar{r}\right] \frac{\delta \rho}{\rho}dr \nonumber \\ 
&-\int_{0}^{R}\frac{Gm \rho}{r^{2}}\left[\int_{0}^{r} K^{n,l}_{A,Y} \frac{Gm}{\bar{r}c^{2}} \left[ \frac{\partial \ln \Gamma_{1}}{\partial \ln P}\vert_{\rho,Y} +1\right]d\bar{r}\right] \frac{\delta \rho}{\rho}dr \nonumber \\
&-\int_{0}^{R} 4 \pi r^{2} \rho \left[ \int_{r}^{R} \frac{G \rho}{\bar{r^{2}}}\left( \int_{0}^{\bar{r}} \frac{G K^{n,l}_{A,Y}}{\tilde{r}c^{2}P} \left( \frac{\partial \ln \Gamma_{1}}{\partial \ln P}\vert_{\rho,Y} +1 \right) d \tilde{r} \right)d\bar{r}\right] \frac{\delta \rho}{\rho}dr \nonumber \\
&-\int_{0}^{R} K^{n,l}_{A,Y}\frac{G m}{r c^{2}} \left[ \frac{\ln \Gamma_{1}}{\ln \rho}\vert_{P,Y} -1 \right] \frac{\delta \rho}{\rho}dr = \int_{0}^{R}K^{n,l}_{\rho,Y} \frac{\delta \rho}{\rho}dr. \label{EqIntegAY}
\end{align}
The first term of this equation is exactly zero for $r=0$. However, it is not for $r=R$ and it is a strong hypothesis to consider that the surface relative density differences are exactly zero. Nevertheless, it is the only way to obtain an equation for the $(A,Y)$ kernels with the direct method. The problem is exactly the same for the $(N^{2},c^{2})$ kernels which can be easily derived but will also face the same problem due to the density derivative. Moreover, one could argue that the contribution of the surface term is negligible when compared to the integrals and that this term has no impact on the final result of the kernels\footnote{This could be done by analysing how accurate Eq. \ref{EqIntegAY} is without the additional term.}. This simplification could also be seen as a boundary condition, stating that the kernels we are searching for have to be exactly $0$ at the surface boundary. Ultimately, after a few additional algebraic operations, we obtain a third order differential equation that we write here as a function of $\mathcal{K}=\frac{K^{n,l}_{A,Y}}{r^{2}\rho}$ and $\mathcal{K}^{'}=\frac{K^{n,l}_{\rho,Y}}{r^{2} \rho}$. 
\begin{align}
&-\frac{r^{6}}{m}\frac{d^{3}\mathcal{K}}{dr^{3}}+\left[ \frac{4 \pi r^{8} \rho}{m^{2}} - \frac{10 r^{5}}{m} -\frac{r^{6}}{m} \frac{d\ln \rho}{dr} - \frac{r^ {5}A_{1}}{m} \right] \frac{d^{2}\mathcal{K}}{dr^{2}} \nonumber \\
& - \left[ \frac{20r^{4}}{m}-\frac{16\pi r^{7}\rho}{m^{2}} + \frac{2r^{6}}{m}\frac{d^{2}\ln \rho}{dr^{2}} +2\frac{r^{5}}{m}\frac{dA_{1}}{dr} +\frac{7r^{5}}{m}\frac{d \ln \rho}{dr} \right. \nonumber \\ & \left.- \frac{4\pi r^{8} \rho}{m^{2}} + \frac{5r^{4}A_{1}}{m}-\frac{4\pi r^{7}\rho A_{1}}{m^{2}}- \frac{4 \pi r^{6} \rho}{c^{2}m}\right] \frac{d \mathcal{K}}{dr} \nonumber \\
&-\left[\left[ \frac{5r^{4}}{m} -\frac{4 \pi r^{7} \rho}{m^{2}}\right]\left[\frac{d \ln \rho}{dr}+\frac{dA_{1}}{dr}\right] + 4G A_{2} \right. \nonumber \\
& \left. +\frac{r^{5}}{m} \frac{d^{2}A_{1}}{dr^{2}} +\frac{r^{6}}{m}\frac{d^{3} \ln \rho}{dr^{3}}+\frac{16 \pi^{2}r^{8}\rho^{2}}{c^{2}m^{2}}+\frac{8 \pi r^{6}\rho}{c^{3}m}\frac{dc}{dr} \right. \nonumber \\ 
& \left. -\frac{4\pi r^{6}}{c^{2}m}\frac{d\rho}{dr}+\frac{24 \pi r^{5}\rho}{c^{2}m}+\left[ \frac{7r^{5}}{m}-\frac{4 \pi r^{8}\rho}{m^{2}}\right] \frac{d^{2} \ln \rho}{dr^{2}}\right] \mathcal{K} \nonumber \\
&= \frac{r^{5}}{m}\frac{d^{2} \mathcal{K}^{'}}{dr^{2}} + \left[\frac{5r^{4}}{m}- \frac{4 \pi r^{7} \rho}{m}\right] \frac{d \mathcal{K}^{'}}{dr}, \label{EqDiffAYdirect}
\end{align}
with the following additional definitions:
\begin{align}
A_{1}&= \frac{Gm}{rc^{2}}\left[\frac{\partial \ln \Gamma_{1}}{\partial \ln \rho}\vert_{P,Y}-1\right] \\
A_{2}&= \frac{Gm \rho r}{c^{2}P}\left[ \frac{\partial \ln \Gamma_{1}}{\partial \ln P}\vert_{\rho,Y} +1\right]
\end{align}
Now the central boundary conditions are found using additional transformations. Typically, we solve the equation using $r^{2}$ as the independent variable for radial position. Taking the limit of the differential equation when $r^{2}$ goes to $0$ and simplifying leads to simple central boundary conditions. Surface conditions are found by stating that Eq. \ref{EqIntegAY} has to be satisfied and that the kernels must be $0$ at the surface. One must also take care of the discontinuous coefficients, meaning that, again, the system must be solved in the radiative and convective regions independently and reconnected using proper continuity conditions. We face the same problem as for the $(u,Y)$ or $(u,\Gamma_{1})$ kernels but can use the same algebraic manipulation to solve the system. Ultimately, we have to solve three discretized equations on two different domains (six systems in total) and reconnect those solutions. Now in addition to the numerical cost of such manipulations, we can see in Eq. \ref{EqDiffAYdirect} third derivatives of the density and second derivatives of the $A_{1}$ function. From numerical experiments not presented here, we have seen that these coefficients contain numerical noise due to the quality of the reference model. The noise can be reduced by increasing the quality of the model and of the finite difference scheme, but smoothing is still necessary to a certain extent. The concern with the smoothing process is that it could in some pathological cases change significantly the form of the kernels. In conclusion, despite the applicability of the method to the $(A,Y)$ kernels and the fact that it uses different hypotheses to obtain structural kernels, we state that for the $(A,Y)$ kernels, the direct method is not well suited in large scale automated studies. Moreover, due to the intrinsic problems of inversions with the adimensional variables mentioned before and the difficulties in determining accurate radii, the $(A,Y)$ pair might well be restricted to solar inversions for which the conjugated functions method is perfectly valid and should be preferred since it leads to simpler equations. 
\end{document}